\documentclass[prb,twocolumn,showpacs,showkeys,superscriptaddress,amsmath,amssymb,floatfix]{revtex4}
\usepackage{graphicx,texdraw}
\usepackage{dcolumn}
\usepackage{bm}

\begin{document}


\title{Relevance of space anisotropy in the critical behavior of
$m$-axial Lifshitz points}

\author{H.\ W. Diehl}
\affiliation{%
Fachbereich Physik, Universit{\"a}t Duisburg-Essen, 45117 Essen,
Germany
}%
\author{M.\ A. Shpot}
\affiliation{%
Fachbereich Physik, Universit{\"a}t Duisburg-Essen, 45117 Essen,
Germany
}%
\affiliation{%
Institute for Condensed Matter Physics, 79011 Lviv, Ukraine
}%
\author{R.\ K.\ P. Zia}
\affiliation{%
Fachbereich Physik, Universit{\"a}t Duisburg-Essen, 45117 Essen,
Germany
}%
\affiliation{%
Department of Physics, Virginia Polytechnic Institute and State
University, Blacksburg, Virginia 24601, USA
}%

\date{\today}

\begin{abstract}
  The critical behavior of $d$-dimensional systems with $n$-component
  order parameter $\bm{\phi}$ is studied at an $m$-axial Lifshitz
  point where a wave-vector instability occurs in an $m$-dimensional
  subspace ${\mathbb R}^m$ ($m{>}1$). Field theoretic renormalization
  group techniques are exploited to examine the effects of terms in
  the Hamiltonian that break the rotational symmetry of the Euclidean
  group ${\mathbb E}(m)$. The framework for considering general
  operators of second order in $\bm{\phi}$ and fourth order in the
  derivatives $\partial_\alpha $ with respect to the Cartesian
  coordinates $x_\alpha $ of ${\mathbb R}^m$ is presented. For the
  specific case of systems with cubic anisotropy, the effects of
  having an additional term,
  $\sum_{\alpha=1}^m(\partial_\alpha^2\bm{\phi})^2$, are investigated
  in an $\epsilon$~expansion about the upper critical dimension
  $d^{*}(m)=4+m/2$. Its associated crossover exponent is computed to
  order $\epsilon^2$ and found to be positive, so that it is a
  \emph{relevant} perturbation on a model isotropic in ${\mathbb
    R}^m$.
\end{abstract}

\pacs{PACS: 05.20.-y, 11.10.Kk, 64.60.Ak, 64.60.Fr}

\keywords{Lifshitz point, critical behavior, field theory}

\maketitle

\section{Introduction}
\label{sec:Intro}

According to the modern theory of critical
phenomena,\cite{Fis83,Fis98} systems can be divided into universality
classes such that the leading singularities of thermodynamic
quantities of all members of a given class are the same. The key
concept here is that the detailed differences between systems of any
given class are ``irrelevant.'' The most prominent and best studied
universality classes are those of the $d$-dimensional $n$-vector
models with short-range (ferromagnetic) interactions, conveniently
represented by the standard $\phi^4$ model with Hamiltonian
\begin{equation}
  \label{eq:hamphi4}
   {\mathcal{H}}=
  {\int}d^dx{\left[\frac{1}{2}\,{(\nabla
 \bm{\phi} )}^2
+\frac{\mathring{\tau}}{2}\,
\bm{\phi}^2+\frac{\mathring{u}}{4!}\,|\bm{\phi} |^4\right]}\;.
\end{equation}
Here $\bm{\phi}=(\phi_a)$ (with $a=1,\ldots,n$) is an $n$-component
order-parameter field, while $\mathring{\tau}$ and ${\mathring{u}}$
are the bare mass and coupling, respectively. To access a critical
point, $\mathring{\tau}$ must be tuned (corresponding to tuning, say,
the temperature of the system $T$) to a special value
$\mathring{\tau}_c$ so that the renormalized $\tau$ (corresponding to
the inverse susceptibility of the system) vanishes. The importance of
this family of models derives from the fact that an enormous variety
of experimentally studied systems belong to these universality
classes. Specifically, irrelevant microscopic details include the
lattice structure, the range of interactions (assumed finite and
ferromagnetic), or pair interactions decaying with a sufficiently
large power of the separation (e.g., van-der-Waals).\cite{Aha76} Of
course, not all microscopic details are irrelevant. Small admixtures
of such interactions can be treated theoretically as ``relevant
perturbations'' to the above universality classes, leading to
predictions of experimentally measurable behavior of crossovers to
other classes.

There are many extensions of the models (\ref{eq:hamphi4}),
associated with systems with more complex types of microscopic
interactions. For example, systems with \emph{competing} interactions
may display a richer variety of behavior. One particular simplified
version is the axial next-nearest neighbor Ising (ANNNI)
model,\cite{FS80,Sel92} in which an antiferromagnetic interaction
between NNN pairs along one of the axes in a simple cubic lattice, in
addition to the usual nearest-neighbor ferromagnetic interactions, is
present. By tuning two (or more) control parameters in such systems,
one can access a Lifshitz point.\cite{HLS75a}

The focus of this paper is the critical properties and universality
classes of $m$-axial Lifshitz
points,\cite{HLS75a,Lifreviews,Hor80,Sel92,Die02} possible in
generalizations of the uniaxial ($m=1$) ANNNI model. To describe
these, we split the Euclidean space ${\mathbb R}^d$ into ${\mathbb
  R}^m\times {\mathbb R}^{\bar{m}}$ with $\bar{m}\equiv d-m$. Let us
label the coordinates in these subspaces as $x_\alpha $, $\alpha
=1,\ldots ,m$, and $x_\beta $, $\beta =m+1,\ldots ,d$, respectively,
and introduce the notations $\partial_\alpha \equiv \partial /\partial
x_\alpha $ and $\partial_\beta \equiv \partial /\partial x_\beta $.
Then the Hamiltonian of these extended models reads
\begin{eqnarray}
  \label{eq:Hamiso}
  {\mathcal{H}}_{\text{iso}}&=&
  {\int}d^dx{\Bigg[\frac{\mathring{\rho}}{2}\,
     \sum_{\alpha=1}^m{(\partial_\alpha\bm{\phi})}^2
    +\frac{1}{2}\,
     \sum_{\beta=m+1}^d{(\partial_\beta\bm{\phi})}^2}
  \nonumber\\&&{\mbox{}
    +\frac{\mathring{\sigma}_1}{2}\,
     \bigg(\sum_{\alpha=1}^m\partial_\alpha^2\bm{\phi} \bigg)^2
    +\frac{\mathring{\tau}}{2}\,
     \bm{\phi}^2+\frac{\mathring{u}}{4!}\,|\bm{\phi}|^4\Bigg]}\;.
\end{eqnarray}
Provided the microscopic aspects (e.g., $d>d_*(m,n)$, the lower
critical dimension here\cite{Die02}) allow the system to be tuned to
Lifshitz points, they occur at critical values
$\mathring{\rho}_{\text{LP}}$ and $\mathring{\tau}_{\text{LP}}$ of
$\mathring{\rho}$ and $\mathring{\tau}$.  Analogous to the above case,
both of the \emph{renormalized} parameters vanish at these points:
$\rho=0$ and $\tau=0$. For $m=0$, the model (\ref{eq:Hamiso}) reduces
to the standard isotropic $\phi^4$ theory of Eq.~(\ref{eq:hamphi4}),
with no Lifshitz points. At the other extreme, $m=d$, the system
displays an isotropic Lifshitz point.

Although the model (\ref{eq:Hamiso}) was introduced more than 25 years
ago, its systematic investigation beyond Landau theory via modern
methods of field-theoretic renormalization group has just
begun.\cite{MC98,MC99,DS00a,SD01,DS01a,DS02} Within the framework of
an $\epsilon$~expansion ($\epsilon \equiv d^*(m)-d$, $d^*(m)$ is the
upper critical dimension $4+m/2$, and $0\le m \le 8$), early studies
were either restricted to special values of $m$ and a subset of
critical exponents\cite{HLS75a,SG78,HB78} or else produced
results\cite{Muk77,HB78} to $O( \epsilon^2 )$ in conflict with those
of Sak and Grest\cite{SG78} and more recent
work.\cite{MC98,MC99,DS00a,SD01,DS01a,DS02} Only recently has it
become possible to master the enormous technical difficulties one
encounters beyond the one-loop approximation. The full two-loop
renormalization group (RG) analysis yielded results for all exponents
(critical, crossover, and correction-to-scaling) to order $\epsilon^2$
for all values of $m$.\cite{DS00a,SD01,DS01a,DS02} An alternative
picture of the Lifshitz point has been advocated by Leite.\cite{Lei03}
This has been critically assessed in Ref.~\onlinecite{DS03}.

Let us also mention some earlier work on modifications of the model
(\ref{eq:Hamiso}). Hornreich \cite{Hor79} investigated the effects of
contributions breaking the $O(n)$ invariance of the Hamiltonian, using
a one-loop approximation. Folk and Moser \cite{FM93} studied Lifshitz
points in systems with short-range and uniaxial dipolar interactions
such as uniaxial ferroelectrics.

The purpose of the present paper is to examine the legitimacy of
taking the fourth-order derivative terms of the Hamiltonian
(\ref{eq:Hamiso}) as \emph{isotropic} in the subspace ${\mathbb R}^m$.
Made essentially for the sake of simplifying the computations, this
assumption of ``$m$-isotropy'' is questionable, since the discrete
lattice symmetries at microscopic scales are unlikely to respect full
rotational invariance. Of course, we must account for these underlying
symmetries at the continuum level when appropriate Hamiltonians are
considered. Now, in the long wavelength limit, isotropy can be
restored by appropriate rescaling of the axes at the level of second
order derivatives. However, there is no such luxury in general, at the
higher orders. Hence, the replacement
\begin{equation}
  \label{eq:anisoder}
  \mathring{\sigma}_1\,\bigg(\sum_{\alpha=1}^m
  \partial_\alpha^2\bm{\phi} \bigg)^2
  \to {\mathcal T}_{\alpha_1\alpha_2\alpha_3\alpha_4}
  (\partial_{\alpha_1}\partial_{\alpha_2}\bm{\phi})\,
  {\partial_{\alpha_3}\partial_{\alpha_4}\bm{\phi}}
\end{equation}
should be made in Eq.~(\ref{eq:Hamiso}), where ${\mathcal T}$ is a
linear combination
\begin{equation}
  \label{eq:T}
  {\mathcal T}_{\alpha_1\alpha_2\alpha_3\alpha_4} =\mathring{\sigma}_i\,
  T^{(i)}_{\alpha_1\alpha_2\alpha_3\alpha_4}
\end{equation}
of tensors $T^{(i)}$ compatible with the symmetry of the microscopic
model considered. Here the summation convention is used: The doubly
occurring index $i$ as well as all pairs of $\alpha$ indices are to be
summed over. In general (the ``$m$-clinic'' case, a generalization of
the familiar triclinic one for $m=3$), there are
\begin{equation}
  \label{eq:K4m}
  n_m= \binom{m+3}{4} \,.
\end{equation}
such tensors.\cite{LL83,rem:elcoeff} Instead of dealing with the $n_m$
coefficients $\mathring{\sigma}_i$, a convenient set are the $n_m-1$
dimensionless coupling constants
\begin{equation}
  \label{eq:widef}
  \mathring{w}_i=\mathring{\sigma}_i/\mathring{\sigma}_1\;,\quad
          i=2,\ldots,n_m
\end{equation}
along with $\mathring{\sigma}_1$.

In this paper, we will focus our attention on a familiar example:
The symmetry associated with an $m$-cube, i.e., cubic anisotropy.
Besides the totally symmetric tensor
\begin{equation}
  \label{eq:S}
  S_{\alpha_1\alpha_2\alpha_3\alpha_4} \equiv \frac{1}{3}\,
   {\left(\delta_{\alpha_1\alpha_2}\,\delta_{\alpha_3\alpha_4}
   +\delta_{\alpha_1\alpha_2}\,\delta_{\alpha_3\alpha_4}
   +\delta_{\alpha_1\alpha_2}\,\delta_{\alpha_3\alpha_4}\right)}
\end{equation}
we have only another one, namely, the cubic
\begin{equation}
  \label{eq:delta}
  \delta_{\alpha_1\alpha_2\alpha_3\alpha_4}
  \equiv\delta_{\alpha_1\alpha_2}\,
  \delta_{\alpha_2\alpha_3}\,\delta_{\alpha_3\alpha_4}\;.
\end{equation}
Thus Eq.~(\ref{eq:T}) reduces to
\begin{equation}
  \label{eq:Tmcube}
  {\mathcal T}_{\alpha_1\alpha_2\alpha_3\alpha_4} =
  \mathring{\sigma}_1\,S_{\alpha_1\alpha_2\alpha_3\alpha_4}
  +\mathring{\sigma}_2\,\delta_{\alpha_1\alpha_2\alpha_3\alpha_4}\;,
\end{equation}
and the Hamiltonian becomes
\begin{equation}
  \label{eq:Hamaniso}
  {\mathcal H}={\mathcal
    H}_{\text{iso}}+\frac{\mathring{\sigma}_2}{2}\int
  d^dx\,\sum_{\alpha=1}^m\big(\partial^2_\alpha\bm{\phi}\big)^2\;.
\end{equation}

Note that this model should represent the universality class of a
simple generalization of the ANNNI model, i.e., from the uniaxial
Ising ($m=n=1$) to the $m$-axial $O(n)$ case.  Specifically, consider
a simple cubic lattice ${\mathbb Z}^d$ with classical $n$-vector spins
$\bm{s}_{\bm{i}}$ of unit length on its sites $\bm{i}$. Assume that
the spins are coupled in an $O(n)$ symmetric fashion, but with
different characteristics within the two subspaces ${\mathbb R}^m$ and
${\mathbb R}^{\bar{m}}$.  In the former, suppose the interactions are
like those in the ANNNI model: nearest-neighbor ferromagnetic (of
strength $J_1>0$) but second neighbor \emph{antiferromagnetic}
(strength $J_2>0$) along each of the $m$ principal lattice directions.
In the complementary subspace, let the interactions be only
nearest-neighbor ferromagnetic (of strength $J_3>0$).  The lattice
Hamiltonian ${\mathcal H}_{\text{lat}}$ is explicitly
\begin{eqnarray}
  \label{eq:Hlat}
  k_{\text{B}}T\,{\mathcal H}_{\text{lat}}&=&
   - J_1
    \sum_{\substack{\langle\bm{i},\bm{j}\rangle\\ \bm{i}-\bm{j}=\pm
        \bm{e}_\alpha}} \bm{s}_i{\cdot}\bm{s}_j
   + J_2
    \sum_{\substack{\langle\bm{i},\bm{j}\rangle\\ \bm{i}-\bm{j}=\pm
       2\bm{e}_\alpha}} \bm{s}_i{\cdot}\bm{s}_j
\nonumber\\&&\mbox{}
   - J_3
    \sum_{\substack{\langle\bm{i},\bm{j}\rangle\\ \bm{i}-\bm{j}=\pm
      \bm{e}_\beta}} \bm{s}_i{\cdot}\bm{s}_j \;.
\end{eqnarray}
Denoting the Fourier transform of $\bm{s}_{\bm{i}}$ by
$\tilde{\bm{s}}_{\bm{q}}$, we recast this expression in Fourier space:
$\left( q_\alpha,q_\beta \right) $. The first line yields a
contribution $\propto \sum_{\alpha =1}^m[J_2\cos (2q_\alpha )-J_1\cos
(q_\alpha )]$ to the coefficient of $|\tilde{\bm{s}}_{\bm{q}}|^2$. The
Lifshitz point can be accessed, at this naive level, by tuning the
$O(q_\alpha^2)$ term to vanish, i.e., $J_1=4J_2$. Meanwhile, the
$O(q_\alpha^4)$ term is precisely of the form of the
$\mathring{\sigma}_2$ term in the coarse-grained Hamiltonian
(\ref{eq:Hamaniso}). Though this procedure does not directly yield a
$\mathring{\sigma}_1$ term, there are two good reasons that such a
term is unavoidable. Firstly, we generalized the ANNNI model in the
simplest possible manner: All further-neighbor interactions couple
only spins along the principal directions.\cite{com:aniso,FH93} Had we
introduced NNN bonds along diagonals (i.e., $\bm{i}-\bm{j}=\pm
\bm{e}_{\alpha_1}\pm \bm{e}_{\alpha_2}$), there would be a
contribution of the form $\sum_{\alpha_1\ne \alpha_2}q_{\alpha
  _1}^2\,q_{\alpha_2}^2\, |\bm{s}_{\bm{q}}|^2$, which involves both
the symmetric tensor (\ref{eq:S}) as well as the cubic one
(\ref{eq:delta}). Secondly, the isotropic coupling will be
automatically generated when short wave-length degrees of freedom are
integrated out, as will be shown in the RG analysis below. Thus, we
expect that a wide class of lattice models similar to (\ref{eq:Hlat})
will fall into the universality class described by the Hamiltonian
(\ref{eq:Hamaniso}).

Unless stated otherwise, we will restrict our attention, for
simplicity, to this case (i.e., (\ref{eq:Tmcube}), tensor with cubic
symmetry) and study only the Hamiltonian (\ref{eq:Hamaniso}). Our goal
is to examine the effects of this type of anisotropy on the
\emph{isotropic} $m$-axial Lifshitz point.  Generalizing a two-loop RG
analysis of the latter case,\cite{DS00a,SD01} we will show that the
cubic anisotropy $\propto \mathring{\sigma}_2$ is a \emph{relevant}
perturbation, at order $\epsilon^2$.

In the next section we present the formal framework for
renormalization of the $m$-anisotropic model with general tensors of
the form (\ref{eq:T}), including the associated RG equations. In
Sec.~\ref{sec:cube}, we specialize to the case (\ref{eq:Hamaniso})
with only a cubic anisotropy. Since the anisotropy of interest appears
in the momenta of a two point vertex function, a two-loop computation
is necessary. More explicit results, to first order in the cubic
anisotropy are provided, so that its effects on the RG flow near the
isotropic fixed point, as well as scaling properties, can be
investigated. For general values of $m$ and $n$, the $\epsilon$
expansion, to $O(\epsilon^2)$, of the associated crossover exponent,
$\varphi_2(n,m,d)$, is obtained in terms of integrals over a single
variable. In Sec.~\ref{sec:estim}, we compute these integrals,
analytically for the special cases of $m=2,6$ and numerically for a
range of other $m$'s. An estimate of $\varphi_2(1,2,3)$ is presented.
Concluding remarks are reserved for Sec.~\ref{sec:concl}. Finally,
there are three appendixes to which some details of our calculations
have been relegated.

\section{Renormalization group analysis}
\label{sec:RGA}
\subsection{General anisotropy: Renormalization and RG equations }
\label{sec:ren}

To renormalize our theory with general ``$m$-anisotropy''
(\ref{eq:T}), we straightforwardly extend the considerations for the
$m$-isotropic model (\ref{eq:Hamiso}) in Ref.~\onlinecite{DS00a}. For
the details in the analysis, we will follow the conventions and
notations of Ref.~\onlinecite{SD01}. Here, we have $n_m$ variables,
$\mathring{\sigma}_i$, whose scaling dimensions vanish at the Gaussian
fixed point $\mathring{\rho}=\mathring{\tau}=\mathring{u}=0$. There
are two consequences: Associated with each of the
$\mathring{\sigma}_i$ is a renormalization factor $Z_{\sigma_i}$.
Further, these quantities, as well as all other renormalization
factors, become functions of $n_m$ dimensionless coupling constants,
namely, the renormalized four-point coupling $u$ and $n_m-1$
renormalized counterparts of the bare variables $\mathring{w}_i$
[Eq.~(\ref{eq:widef})]:
\begin{equation}
  \label{eq:wi's}
w_i\equiv \sigma_i/\sigma_1\;,\quad i=2,\ldots ,n_m\;.
\end{equation}

Accordingly, we reparametrize the theory as
\begin{eqnarray}
   \mathring{\sigma}_i&=&Z_{\sigma_i}(u,\bm{w})\,\sigma_i\;,\\
   \bm{\phi}&=&[Z_\phi(u,\bm{w})]^{1/2}\,\bm{\phi}_{\rm ren}\;,\\
  \label{eq:rhorep}
   \left(\mathring{\rho}-\mathring{\rho}_{\text{LP}}\right)\,
     {\mathring{\sigma}_1}^{-1/2}
        &=&\mu\,Z_\rho(u,\bm{w})\,\rho\;,\\
 \label{eq:reptau}
  \mathring{\tau}-\mathring{\tau}_{\text{LP}}
        &=&\mu^2\,Z_\tau(u,\bm{w})\,{\big[\tau
          +A_\tau(u,\bm{w})\,\rho^2\big]},\quad\\
 \label{eq:repuw}
  \mathring{u}\,{\mathring{\sigma_1}}^{-m/4}\,F_{m,\epsilon}
        &=&\mu^\epsilon\,Z_u(u,\bm{w})\,u\;,
\end{eqnarray}
where $\mu $ is a momentum scale, $F_{m,\epsilon }$ denotes the
normalization factor
\begin{equation}
  \label{eq:Fmeps}
F_{m,\epsilon}=
\frac{\Gamma{\left(1+{\epsilon/ 2}\right)}
\,\Gamma^2{\left(1-{\epsilon/ 2}\right)}\,
\Gamma{\left({m}/{4}\right)}}{(4\,\pi)^{({8+m-2\,\epsilon})/{4}}\,
\Gamma(2-\epsilon)\,\Gamma{\left({m}/{2}\right)}}\;,
\end{equation}
and $\bm{w}$ stands for the set $\{w_2,\ldots ,w_{n_m}\}$ of $n_m-1$
variables. Following Ref.~\onlinecite{DGR03}, we have included a
renormalization function $A_\tau(u,\bm{w})$ to absorb
momentum-independent poles proportional $\rho^2$ of the two-point
vertex function.

The fact that the theory must reduce for $\bm{w}=\bm{0}$ to the
$m$-isotropic one implies the relations
\begin{eqnarray}
Z_\iota(u,\bm{w}=\bm{0})&=&Z^{\text{SD}}_\iota(u)\;,\quad
  \iota=\phi,u,\tau,\rho\;,\\
Z_{\sigma_1}(u,\bm{w}=\bm{0})&=&Z^{\text{SD}}_\sigma(u)\;,
\end{eqnarray}
where the $Z$~factors marked by the superscript ``SD'' are those of
Ref.~\onlinecite{SD01}. The function $A_\tau(u,\bm{w}=\bm{0})$ has been
computed to one-loop order in Ref.~\onlinecite{DGR03}. Its explicit
form will not be needed in the sequel.

Turning to the RG equations, we use the notation $\partial_\mu |_0$
for $\mu$~derivatives at fixed bare variables ($\mathring{u}$,
$\mathring{\sigma}_i$, $\mathring{\tau}$ and $\mathring{\rho})$, and
define the $\beta $ and exponent functions
\begin{equation}
  \label{eq:betafundef}
  \beta_\kappa\equiv\left.\mu\partial_\mu\right|_0\kappa\;,\quad
  \kappa=u,\tau,\rho,\sigma_i\;,
\end{equation}
\begin{eqnarray}\label{etadef}
\eta_\lambda(u,\bm{w})\equiv\left.\mu\, \partial_\mu\right|_0\ln
Z_\lambda\;,
\quad\lambda=\phi,\,u,\,\tau,\,\rho,\,\sigma_i\;.
\end{eqnarray}
The functions $\eta_\lambda $ depend only on $u$ and $\bm{w}$.  Since we
use minimal subtraction of poles, they are even independent of
$\epsilon $.  In terms of these variables and\cite{DGR03}
\begin{equation}
  \label{eq:btau}
  b_\tau(u,\bm{w})\equiv A_\tau\, {\big[
    \left.\mu\partial_\mu\right|_0\ln A_\tau+
    \eta_\tau-2\eta_\rho\big]}\;,
\end{equation}
the $\beta$~functions can be written as
\begin{eqnarray}\label{eq:betadef}
\beta_u&=&-{\left[\epsilon+\eta_u(u,\bm{w})\right]}\,u\;,\\
\beta_\tau&=&-[2+\eta_\tau(u,\bm{w})]\,\tau
-\rho^2\,b_\tau(u,\bm{w})\;,\label{eq:betataures}\\
\beta_\rho&=&-[1+\eta_\rho(u,\bm{w})]\,\rho\;,\\
 \label{eq:betaSigma}
\beta_{\sigma_i}&=&-\eta_{\sigma_i}(u,\bm{w})\,\sigma_i\;.
\end{eqnarray}
Also, since we fixed all renormalization factors $Z_\lambda$ such that
the regular part of their Laurent series in $\epsilon$ is exactly
unity, the associated $\eta_\lambda$ functions are related to the
residues of the $Z_\lambda$'s via
\begin{equation}
  \label{eq:etaminsub}
  \eta_\lambda=-u\,\partial_u\mathop{\text{Res}}_{\epsilon=0}Z_\lambda\;.
\end{equation}

In terms of the operator
\begin{equation}
{\mathcal{D}}_\mu \equiv \mu \,\partial_\mu +\sum_\kappa \beta_\kappa
\,\partial_\kappa \;,
\end{equation}
the RG~equations of the renormalized $N$-point cumulant functions
$G_{\text{ren}}^{(N)}(\bm{x})\equiv\big\langle
\prod_{j=1}^N\phi_{a_j,{\rm ren}}(\bm{x}_j)\big\rangle^{\text{cum}}$
and the corresponding vertex functions $\Gamma_{\text{ren}}^{(N)}$
read, respectively,
\begin{equation}
\left[ {\mathcal{D}}_\mu +\frac N2\,\eta_\phi \right] G_{\text{ren}%
}^{(N)}=0\;,\quad \left[ {\mathcal{D}}_\mu -\frac N2\,\eta_\phi \right]
\Gamma_{\text{ren}}^{(N)}=0\;.  \label{eq:RGG}
\end{equation}

Being dimensionless, $\bm{w}$ will appear in the $Z$'s to
arbitrary orders in general, even though we are dealing with the
systematics of an expansion in powers of $u$ (the loop expansion).
However, our principal goal here is a local stability analysis of the
model (\ref{eq:Hamaniso}) about the isotropic fixed point
\begin{equation}
\mathcal{P}_{\text{iso}}^{*} :  (u^*,\bm{w}=\bm{0})
\label{eq:isoFP}
\end{equation}
where $u^{*}$ is the nontrivial zero of $\beta_u(\epsilon
,u,\bm{w}=\bm{0})$ the explicit form of which, up to $O(\epsilon
^2)$, is given in Eq.~(60) of Ref.~\onlinecite{SD01}. To this end, we
can linearize about $\mathcal{P}_{\text{iso}}^{*}$. Hence it will be
sufficient to compute the counterterms to first order in $\bm{w}$.

\subsection{Cubic anisotropy: RG flow and scaling}
\label{sec:cube}

Given the general framework above, we turn to explicit results for the
particular case with only a cubic anisotropy. With just $\sigma_1$ and
$\sigma_2$, we have only one $w$ (and no need for the set
$\bm{w}$).  Further, we need only terms linear in $w$. Referring
the reader to Appendix~\ref{sec:appGamma} for the computational
details, we note here that the pole terms of the two-loop graph
\raisebox{-4pt}{\begin{texdraw} \drawdim pt \setunitscale 2.2 \linewd
    0.3 \move(-3 0) \move(5 0) \lellip rx:5 ry:2.5 \move(-2 0)
\rlvec(14 0) \move(13 0)
\end{texdraw}}
yield the renormalization factors
\begin{eqnarray}\label{eq:Zphi}
Z_\phi&=&1-\frac{n+2}{3}\,\frac{1}{12\,(8-m)}
\\\nonumber&&\mbox{} \times
\Big[j_\phi(m)-36\,i_\phi(m)\,w \Big]\frac{u^2}{\epsilon}+O(w^2,u^3)\;,
\end{eqnarray}
\begin{eqnarray}
  \label{eq:ZphiZsig1}
  Z_\phi Z_{\sigma_1}&=&1+\frac{n+2}{3}\,\frac{1}{96\,m(m+2)}
\\\nonumber&&\mbox{} \times
\Big[j_\sigma(m)-36\, i_{\sigma_1}(m)\,w \Big]\frac{u^2}{\epsilon}
   +O(w^2,u^3)\;,
\end{eqnarray}
and
\begin{eqnarray}
  \label{eq:ZphiZsig2}
  Z_\phi Z_{\sigma_2}&=&1-\frac{n{+}2}{3}\,
  \frac{4\,i_{\sigma_2}(m)}{m(m+2)(m+4)(m+6)}\frac{u^2}{\epsilon}
\nonumber\\&&\mbox{}
+O(wu^2,u^3)\;.
\end{eqnarray}
Here, $j_\phi (m)$ and $j_\sigma (m)$ are single-variable integrals
encountered in Ref.~\onlinecite{SD01}, their definitions noted for the
readers' convenience in Appendix~\ref{sec:appGamma}:
Eqs.~(\ref{eq:jphidef}) and (\ref{eq:jsigmadef}). Though more
complicated, $i_\phi (m)$, $i_{\sigma_1}(m)$, and $i_{\sigma_2}(m)$
are analogous integrals, defined in
Eqs.~(\ref{eq:iphidef})--(\ref{eq:isigma2def}).

From the structure of these $Z$ factors, it is clear that, at this
two-loop order, both $\eta_{\sigma_1}$ and $\eta_{\sigma_2}$ are
proportional to $\frac{n+2}3u^2$ with coefficients linear in $w$. As a
result of Eq.~(\ref{eq:betaSigma}), and keeping only terms \emph{to
  first order} in $\sigma_2$, we find the associated $\beta $
functions to be of the form
\begin{equation}
\left(
\begin{array}{c}
\beta_{\sigma_1} \\
\beta_{\sigma_2}
\end{array}
\right) =-2\frac{n+2}3u^2\left(
\begin{array}{cc}
K_{11} & K_{12} \\
0 & K_{22}
\end{array}
\right) \left(
\begin{array}{c}
\sigma_1 \\
\sigma_2
\end{array}
\right) [1+O(w)]\;.  \label{eq:betasiglin}
\end{equation}
where
\begin{eqnarray*}
K_{11} &\equiv &\frac{j_\phi (m)}{12\,(8-m)}
    +\frac{j_\sigma (m)}{96\,m(m+2)} \;, \\
K_{12} &\equiv &-\frac{3\,i_\phi (m)}{8-m}
    -\frac{3\,i_{\sigma_1}(m)}{8m(m+2)} \;, \\
K_{22} &\equiv &\frac{j_\phi (m)}{12\,(8-m)}
    -\frac{4\,i_{\sigma_2}(m)}{m(m+2)(m+4)(m+6)} \;,
\end{eqnarray*}
are constants, independent of the couplings.

Since our main interest is the neighborhood of
$\mathcal{P}_{\text{iso}}^{*}$, we need to evaluate $u$ in this
equation only at the $m$-isotropic fixed point
\begin{equation}
u^{*}=\frac{6\epsilon }{n+8}+O(\epsilon^2).
\end{equation}
Given the form of the matrix in Eq.~(\ref{eq:betasiglin}), the
eigenvalues are trivially obtained, the first of which is just $\eta
_\sigma^{*}$ in Refs.~\onlinecite{DS00a} and \onlinecite{SD01}:
\begin{eqnarray*}
\eta_{\sigma_1}^{*} &\equiv &\eta_\sigma^{*}
   =-\frac{24\,(n+2)}{(n+8)^2}\,K_{11}\,\epsilon^2+O(\epsilon^3 )\;,\\
\eta_{\sigma_2}^{*} &\equiv &
   -\frac{24\,(n+2)}{(n+8)^2}\,K_{22} \, \epsilon^2+O(\epsilon^3 )\;.
\end{eqnarray*}
Associated with these are, respectively, the (linear) scaling fields:
\begin{equation}
\sigma_1+b\,\sigma_2\quad \text{and}\quad \sigma_2\;,
\end{equation}
where $b\equiv {K_{12}}/{K_{11}-K_{22}}$.  Near
$\mathcal{P}_{\text{iso}}^{*}$, we may drop the irrelevant
contributions proportional to $u-u^{*}$, so that the flow equations
\begin{equation}
\ell \frac \partial {d\ell }\bar{\sigma}_i=\beta_{\sigma_i}
\label{eq:fleqsigmas}
\end{equation}
are solved by $\bar{\sigma}_i(\ell)$. Imposing the initial conditions
$\bar{\sigma}_i(1) = \sigma_i$, these take the asymptotic forms
\begin{eqnarray}
\bar{\sigma}_1(\ell )&\approx&
   (\sigma_1+b\,\sigma_2)\,\ell^{-\eta_\sigma^{*}}
    -b\,\sigma_2\,\ell^{-\eta_{\sigma_2}^{*}}\;,
\label{eq:linflsigmas} \\
    \bar{\sigma}_2(\ell ) &\approx & \sigma_2\,
    \ell^{-\eta_{\sigma_2}^{*}}\;.
\end{eqnarray}
Thus, to the order of interest, the anisotropy $w$ leads to a
dependence on the following ratio of running variables:
\begin{eqnarray}
\bar{\sigma}_2(\ell )/\bar{\sigma}_1(\ell )&\approx&
(\sigma_2/\sigma_1)\,\ell^{-(\eta_{\sigma_2}^{*}-
  \eta_\sigma^{*})}\\
&=& {\big[w+O(w^2)\big]}\, \ell^{-(\eta_{\sigma_2}^{*}
  -\eta_\sigma^{*})}\;.
\label{eq:anisscf}
\end{eqnarray}

As we will show, ${\eta_{\sigma_2}^{*}-\eta_\sigma^{*}}$ is
\emph{positive}, so that the effect of $w$ is more significant in the
infrared limit ($\ell \rightarrow 0$). Therefore, we introduce the
anisotropy crossover exponent
\begin{equation}
\varphi_2\equiv \nu_{l2}\,(\eta_{\sigma_2}^{*}-\eta_\sigma^{*})\;,
\label{eq:varphi2def}
\end{equation}
which governs the scaling behavior of $w$ with $\tau $: $w\sim
\tau^{-\varphi_2}$. Since the $\eta $'s are already of
$O(\epsilon^2)$, we may insert the zeroth-order value for $\nu_{l2}$
(i.e., $1/2$) to obtain
\begin{eqnarray}
  \label{eq:phi2}
  \varphi_2&=&\frac{n+2}{(n+8)^2}\,\frac{1}{m(m+2)}
   \nonumber \\[-1em]&&\\[-0.5em]&&\nonumber \times
   {\bigg[\frac{48\,i_{\sigma_2}(m)}{(m+4)(m+6)}
   +\frac{j_\sigma(m)}{8}\bigg]}\,\epsilon^2+O(\epsilon^3)\;.
\end{eqnarray}

As a consequence of the contribution proportional to $\rho^2$ of
$\beta_\tau$ [see Eq.~(\ref{eq:betataures})], the variable $\tau$ is
not a scaling field. Proceeding similarly as in
Ref.~\onlinecite{DGR03}, we can define a nonlinear scaling
field\cite{Weg72a}
\begin{equation}
  \label{eq:gtaudef}
  g_\tau=\tau+c^\tau_{\rho^2}(u)\,\rho^2+c^\tau_{\rho^2,w}(u)\,w\,\rho^2
+\ldots 
\end{equation}
with the asymptotic scale dependence $\bar{g}_\tau(\ell)\sim
l^{-1/\nu_{l2}}\, g_\tau$, where the ellipsis stands for terms of higher order in $w$.
 
Utilizing the above results, one can generalize the considerations of
Refs.~\onlinecite{DS00a} and \onlinecite{DGR03} in a straightforward
fashion to obtain the scaling forms of the renormalized $N$-point
cumulants $G^{(N)}$:
\begin{eqnarray}
\lefteqn{G^{(N)}[\{x_\alpha ,x_\beta \};g_\tau ,\rho ,\sigma_1,\sigma
_2,u,\mu )]}  \nonumber  \label{eq:scfG} \\
&\approx &g_\tau^{-\nu_{l2}\,\Delta_G}\, \Upsilon_G{\Big[{\Big\{
    \frac{\sqrt{\mu}\,x_\alpha}{\sigma^{1/4}\,g_\tau^{\nu_{l4}}},
    \frac{\mu \,x_\beta}{g_\tau^{\nu_{l2}}} \Big\}};
  \frac{\rho}{g_\tau^\varphi}, \frac{w}{g_\tau^{\varphi_2}}\Big]}\,,\;
\end{eqnarray}
where $\Delta_G$ is the scaling dimension of $G^{(N)}$:
\begin{equation}
  \label{eq:scdim}
\Delta_G=(N/2)\,[d-2+\eta_{l2}+m\,(\theta -1)]\;.
\end{equation}
With the exception of $\varphi_2$, explicit expressions for all exponents,
up to $O(\epsilon^2)$, are given in Refs.~\onlinecite{DS00a} and
\onlinecite{SD01}. In the next section, we present a brief summary of the
steps for computing $\varphi_2$.

\section{Explicit results for the anisotropy crossover exponent}
\label{sec:estim}

From Eq.~(\ref{eq:phi2}), we see that only two integrals are needed
for finding the exponent $\varphi_2$, namely, $j_\sigma (m)$ and
$i_{\sigma_2}(m)$. As shown in Appendix~\ref{sec:appGamma}, all
integrations can be performed analytically, except the one over the
scaling variable
\begin{equation}
\upsilon \equiv \frac{\sqrt{x_\alpha x_\alpha }}{(x_\beta x_\beta)^{1/4}}\;,
\end{equation}
for which we resort to numerical means. The first, $j_\sigma $, is
familiar from Ref.~\onlinecite{SD01}, recalled here in
Table~\ref{tab:jis}.

The second involves, in general, a product of
three hypergeometric functions. As outlined in Appendix E of
Ref.~\onlinecite{SD01}, we write
\begin{eqnarray}
i_{\sigma_2}(m) &=&{\int_0^\infty }d\upsilon \,I_m(\upsilon )  \nonumber
\label{eq:iint} \\
&=&{\int_0^{\upsilon_0}}d\upsilon \,I_m(\upsilon )+R_m(\upsilon_0)\;,
\end{eqnarray}
splitting the integral into a contribution from a finite interval
$(0,\upsilon_0)$ and a remainder $R_m(\upsilon_0)$. The point
$\upsilon_0$ was chosen so that standard numerical integration
routines (specifically, {\sc Mathematica}\cite{rem:MAT}) yield
sufficiently accurate results for the first term \emph{and} a few
terms of the asymptotic expansion for $I_m$ suffice for evaluating
$R_m$ approximately. In practice, we chose $\upsilon_0\simeq 9.5$ and
just the leading term of the asymptotic expansion. The latter can be
computed analytically and leads to:
\begin{equation}
  \label{eq:Rm}
  R_m(\upsilon_0)\approx \frac{3\,\sqrt{\pi }\;2^{8 - m}(m-2)^2}
    {\Gamma{\big(\frac{m}{2}\big)}\,
      \Gamma{\big(\frac{1}{2}+\frac{m}{4}\big)
    \,\Gamma{\big(2-\frac{m}{4}\big)}}}\,
 \,\frac{\upsilon_0^{m-8}}{8-m}\;.
\end{equation}
Combining this with the numerical integration over $(0,\upsilon_0)$, we
arrive at the values of $i_{\sigma_2}(m)$ displayed in
Table~\ref{tab:jis}.\cite{rem:accuracy}

As discussed in Ref.~\onlinecite{SD01}, the cases $m=2$ and $m=6$ are
quite special. The scaling functions from which $I_m(\upsilon )$ is
formed --- and hence $I_m(\upsilon )$ itself --- reduce to elementary
functions. (From another perspective, their asymptotic expansions
terminate at low orders:\cite{DS00a,SD01} See Eqs.~(\ref{fidva}),
(\ref{eq:Phidstar6x}) and (\ref{eq:Yiv2}), (\ref{eq:Yiv6}).
Indeed, the approximation (\ref{eq:Rm}) even
vanishes for $m=2$.) As a result, $i_{\sigma_2}$ can be computed
analytically. The results (see Appendix~\ref{sec:m2m6}),
\begin{equation}
i_{\sigma_2}(2)=\frac 2{27}\;,\quad i_{\sigma_2}(6)=20\left( 8\,\ln \frac
43-\frac{59}{27}\right) \;,  \label{eq:isigma2m26}
\end{equation}
are in conformity with the values quoted in Table~\ref{tab:jis} and
provide useful checks of our numerical procedure. For completeness,
let us also recall the analytic values
\begin{equation}
\label{eq:jsigmam26}
j_\sigma (2)=\frac{128}{27}\;,\quad j_\sigma (6)=\frac{448}{9}\;
\end{equation}
from Ref.~\onlinecite{SD01}.

Inserting the values from Table~\ref{tab:jis} into
Eq.~(\ref{eq:phi2}), we write the anisotropy crossover exponent as
\begin{equation}
\label{eq:varphi2epsexp}
\varphi_2=\frac{27\,(n+2)}{(n+8)^2}\,C(m)\,\epsilon^2+O(\epsilon^3)\;.
\end{equation}
Here the $C(m)$ are the $\epsilon^2$~coefficients of the crossover
exponent $\varphi_2$ for $n=1$. They are listed in the last column.

For the special cases where analytic results are available, we have
\begin{equation}
  \label{eq:varphi2m2}
  C(2) =\frac{1}{324}
\end{equation}
and
\begin{equation}
  \label{eq:varphi2m6}
   C(6) =\frac{1}{27}\,
     {\bigg[\frac{4}{3}\ln \frac 43-\frac{19}{81}\bigg]}\;.
\end{equation}
As we see, in all cases, $\varphi_2$ is positive at this order. Thus,
we conclude that the isotropic fixed point
${\mathcal{P}}_{\text{iso}}^{*}$ is unstable against perturbations
from cubic anisotropy of the form included in the Hamiltonian
(\ref{eq:Hamaniso}).

\begin{table}[htb]
\caption{Numerical values of the integrals $j_\sigma(m)$,
         $i_{\sigma_2}(m)$, and the coefficients
         $C(m)$ of $\varphi_2$ introduced in
         Eq.~(\ref{eq:varphi2epsexp}).\label{tab:jis}}
\begin{ruledtabular}
\begin{tabular}{cllll}
$m$ &$j_\sigma(m)$&$ i_{\sigma_2}(m)$&$C(m)$\\\hline
2&\phantom{0}4.74074&0.074074&0.00309\\
3& 10.804&0.24682&0.00380\\
4&20.067&0.6175&0.00444\\
5&32.95&1.279&0.00501\\
6&49.7778&2.325428&0.00552
\end{tabular}
\end{ruledtabular}
\end{table}

Of specific interest is the scalar biaxial case ${m=2}$, ${n=1}$,
corresponding to a biaxial generalization of the three-dimensional
ANNNI model discussed in the Introduction. The upper critical
dimension being $5$, let us set $\epsilon =2$ to consider a physical
system in $d=3$:
\begin{equation}
\varphi_2(n{=}1,m{=}2,d{=}3)\simeq \frac 1{81}\simeq 0.0123\;.
\label{eq:varphi2m2num}
\end{equation}

Since this exponent is so small, we see that, unless the anisotropic
amplitudes are large, we must be extremely close to the Lifshitz point
(by careful tuning of the two control parameters) in order to detect
any serious deviations from the critical behavior in the class of the
isotropic fixed point ${\mathcal P}_{\text{iso}}^{*}$.

\section{Concluding remarks}
\label{sec:concl}

We studied the critical properties of $m$-axial Lifshitz points in
systems with spatial anisotropic interactions. Specially, for an
$m$-dimensional subspace of ${\mathbb R}^d$ in which a wave-vector
instability occurs, we considered the effects of arbitrary
fourth-order couplings of the form
\begin{equation}
\label{eq:cub}
{\mathcal T}_{\alpha_1\alpha_2\alpha_3\alpha_4}\,
   q_{\alpha_1}q_{\alpha_2}q_{\alpha_3}q_{\alpha_4}\,
   {\bm{\phi}}_{\bm{q}}\cdot{\bm{\phi}}_{-\bm{q}}\;,
\end{equation}
with $\bm{\phi}$ being an $n$-component order parameter. Unlike
previous studies of ``anisotropic'' Lifshitz points,\cite{Hor79}
we are not concerned with couplings that break the $O(n)$ symmetry
of the order parameter. In
this sense, our $\bm{\phi}$ is more appropriate for, say,
multi-component alloys than for spin systems. Especially for the $n=1$
(Ising) case, we showed how anisotropy of the form (\ref{eq:cub})
naturally arises from a generalized ANNNI model. In the case we
explicitly considered only a cubic anisotropy, corresponding to a
tensor (\ref{eq:delta}), was present [cf.\ Eqs.~(\ref{eq:Tmcube}) and
(\ref{eq:Hamaniso})].

Using field-theoretic renormalization group techniques and expanding
about the upper critical dimension, we found that this is a
\emph{relevant} perturbation for an isotropic $m$-axial Lifshitz
point. To $O(\epsilon^2)$, the crossover exponent has been computed
for a range of $m$, with analytic forms for the special cases of
$m=2,6 $. Though we have not obtained similar results for general
anisotropic interactions, there is no doubt that cubic anisotropy of
the form included in the Hamiltonian (\ref{eq:Hamaniso}) will be
generically present even in systems with lower symmetries, leading to
crossover in general. Since all previous investigations\cite{ML78}
of $m$-axial
Lifshitz points (with $m>1$) are based on $m$-isotropic Hamiltonians,
our conclusion is that, unless the microscopics of a system enforces
rotational invariance in ${\mathbb R}^m$, the critical properties will
not fall into the universality classes found so far. Nevertheless, we
should caution that, though the crossover exponent is positive, its
numerical values are relatively small. For example, in the case most
likely to be physically accessible ($d=3,m=2,n=1$), this exponent is
only $1/81$.
Though measurable deviations from the isotropic class may be difficult
to detect for real systems, it should be interesting to test our predictions
in Monte Carlo simulations of suitably designed lattice models.

A more interesting question is, given the RG flow is away from the
$m$-isotropic fixed point, whether there is a new stable fixed point
or not? Of course, if the former is true, then we have a new
universality class. However, preliminary studies indicate that the
$\beta$ function for $w$ has no zero, for any finite, non-vanishing $w$.
Worse, the flow seems to run indefinitely towards Hamiltonians associated
with singular propagators (i.e., parts quadratic in $\bm{\phi}$ being no
longer positive definite), so that
$O(\bm{q}^6)$ terms
will be needed to stabilize the theory. In addition, we might speculate
that the system would undergo a first order transition, albeit a weak
one. Clearly, further investigations are necessary before definitive
conclusions can be reached.

\begin{acknowledgments}
  We gratefully acknowledge the financial support of this work by the
  Alexander von Humboldt Foundation, the US National Science
  Foundation (through DMR-0088451), and by the Deutsche
  Forschungsgemeinschaft (DFG) --- in its initial phase via the
  Leibniz program (grant \# Di 378/2) and Sonderforschungsbereich 237,
  and in its final phase via DFG grant \# Di-378/3. RKPZ and MASh
  thank Fachbereich Physik for their hospitality at the University of
  Duisburg-Essen.
\end{acknowledgments}

\appendix

\section{Calculation of the two-loop graph of $\Gamma^{(2)}$ to order
  $\mathring{\sigma}_2$} \label{sec:appGamma}

Let us denote momenta in ${\mathbb R}^d$ by $\bm{q}=(\bm{k},\bm{p})$
with components $\bm{k}\in {\mathbb R}^m$ and $\bm{p}\in {\mathbb
  R}^{\bar{m}}$ (${\bar{m}}\equiv d-m$). As we will be dealing with
cubic spatial anisotropy extensively, let us define
\begin{equation}
\check{k}^4\equiv \sum_{\alpha =1}^mk_\alpha^4\;,
\end{equation}
which is to be distinguished from the usual $k^4\equiv (
\bm{k}{\cdot}\bm{k})^2$. At the Lifshitz point $\tau =\rho =0$, the
Fourier transform of $G(\bm{x})$, the free propagator in position
space, is
\begin{equation}
\tilde{G}(\bm{q})\equiv \tilde{G}(\bm{k},\bm{p})={\left[ p^2 +
\mathring{\sigma}_1\,k^4+\mathring{\sigma}_2\,\check{k}^4\right] }^{-1}\,.
\end{equation}
To two loops, $\tilde{\Gamma}^{(2)}(\bm{q})$, the Fourier transform of
the bare vertex function $\Gamma^{(2)}$, is given by
\begin{equation}
  \label{eq:Gamma2}
\tilde{\Gamma}^{(2)}(\bm{q})=\frac 1{\tilde{G}(\bm{q})}-\frac{n+2}{3}\,
 \frac{\mathring{u}^2}{6}\,\widetilde{G^3}(\bm{q})+O(\mathring{u}^3)\;,
\end{equation}
where $\widetilde{G^3}(\bm{q})$ denotes the Fourier transform of
$G^3(\bm{x})$.  In addition, at this order of the $\epsilon
$~expansion, we will need only the residues
of the simple pole of $\tilde{\Gamma}^{(2)}$
at $\epsilon =0$. Thus, all coefficients of
$\epsilon^{-1}$ may be evaluated at $\epsilon =0$, or $d=d^*=4+m/2$.

Let us compute the pole terms of $ \widetilde{G^3}$ to first order in
$\mathring{w}\equiv\mathring{\sigma}_2/\mathring{\sigma}_1$. Since the
dependence on $\mathring{\sigma}_1$ follows from dimensional
considerations, we temporarily set $\mathring{\sigma}_1=1$. Expanding
in $\mathring{w}$, we have
\begin{equation}
  \label{eq:G3poles}
\widetilde{G^3}={\left[ \widetilde{G^3}\right] }_{\mathring{w}=0} +
\mathring{w}\,
{\left[ \partial_{\mathring{w}}\widetilde{G^3}\right] }_{\mathring{w}=0}
+ O(\mathring{w}^2)\;.
\end{equation}
The pole contribution of the first term on the right-hand side has
been computed in Ref.~\onlinecite{DS00a}. Expressed in terms of the
integrals $j_\phi(m)$ and $j_\sigma(m)$ of Ref.~\onlinecite{SD01} [see
its equations (43), (44), and (46)], the result is
\begin{eqnarray}
  \label{eq:G3wzero}
{\left[ \widetilde{G^3}\right]}_{\mathring{w}=0}\, &=&
\frac{F_{m,\epsilon}^2}{\epsilon}
\left[ -\frac{j_\phi (m)}{2\,(8-m)}\,p^2+
  \frac{j_\sigma(m)}{16\,m(m+2)}k^4\right]
\nonumber\\&&\mbox{} + O(\epsilon^0)\;.
\end{eqnarray}
To compute the $\mathring{w}$ derivative appearing in
Eq.~(\ref{eq:G3poles}), we start from its position-space
representation, $G^3(\bm{x})$, and obtain
\begin{equation}
  \label{eq:G3linw}
  {\left[\partial_{\mathring{w}}G^3(\bm{x})\right]}_{\mathring{w}=0}
= -3\,G_0^2(\bm{x})\sum_{\alpha}\partial_\alpha^4(G_0*G_0)(\bm{x})\;.
\end{equation}
Here $G_0(\bm{x})\equiv G(\bm{x})|_{\mathring{w}=0}$, and
$(G_0*G_0)(\bm{x})$ means a convolution in position space (i.e.,
Fourier transform of $[\tilde{G}_0(\bm{q})]^2$).

Next, let us exploit the scaling forms of both $G_0$ and $G_0*G_0$.
Defining the radii
\begin{equation}
  \label{eq:rR}
  r\equiv\sqrt{x_\alpha\,x_\alpha}\;,\quad R\equiv
  \sqrt{x_\beta\,x_\beta}\;,
\end{equation}
the scaling variables
\begin{equation}
  \label{eq:scalvar}
\upsilon_\alpha \equiv x_\alpha \, R^{-1/2}\;,\quad
z \equiv \upsilon_\alpha \, \upsilon_\alpha = \upsilon^2 \;,
\end{equation}
and the vectors $\bm{\upsilon} \equiv \{\upsilon_\alpha \}$ and
$\bm{e} \equiv \{x_\beta \}/R$, we write
\begin{equation}
  \label{eq:freeprop}
  G_0(\bm{x})=R^{-2+\epsilon}\,\Phi_{m,d}(\upsilon)
\end{equation}
and
\begin{equation}
  \label{eq:scfGG}
  (G_0*G_0)(\bm{x})=R^{\epsilon}\,Y_{m,d}(z) \;.
\end{equation}
Here, the scaling functions are
\begin{equation}
  \label{eq:Fpsil}
  \Phi_{m,d}(\upsilon)\equiv
  \int^{(m)}_{\bm{k}}\int^{(\bar{m})}_{\bm{p}}
  \frac{e^{i\bm{k}{\cdot}\bm{\upsilon}+i\bm{p}{\cdot}\bm{e}}}{k^4+p^2}
\end{equation}
and
\begin{equation}
  \label{eq:Ypsil}
  Y_{m,d}{\big(\upsilon^2\big)}\equiv
  \int^{(m)}_{\bm{k}}\int^{(\bar{m})}_{\bm{p}}
  \frac{e^{i\bm{k}{\cdot}
      \bm{\upsilon}+i\bm{p}{\cdot}\bm{e}}}{(k^4+p^2)^2}
\end{equation}
where $\int_{\bm{k}}^{(m)}\equiv (2\pi)^{-m}\int d^m{k}$ and
$\int_{\bm{p}}^{(\bar{m})}\equiv (2\pi)^{-\bar{m}}\int
d^{\bar{m}}{p}$.  Their explicit expressions in terms of Taylor series
and hypergeometric functions, as well as the asymptotic expansions for
large $z$ are given in Appendix~\ref{sec:scfY}.

Inserting the above scaling forms into Eq.~(\ref{eq:G3linw}) and
performing the Fourier transformation, we encounter integrations over
$\bm{x}$. Using hyperspherical coordinates $(\upsilon
\sqrt{R},\Omega_m)$ in ${\mathbb R}^m$ and $(R,\Omega_{\bar{m}})$ in
${\mathbb R}^{\bar{m}}$, let us denote angular averages by
\begin{equation}
  \label{eq:angav}
  \overline{f}^{D}\equiv S_D^{-1} {\int}f(\Omega_D)\,d\Omega_D
\end{equation}
where $ S_D\equiv\int d\Omega_D=2\pi^{D/2}/{\Gamma(D/2)}$ is the
surface area of a $D$-dimensional unit sphere.  In the radial
integration ${\int_0^\infty}dR$, the distribution $R^{-3+2\epsilon}$
is found to appear. Employing its Laurent expansion\cite{GS64}
\begin{equation}
  \label{eq:epsexprpow}
  R^{-3+2\epsilon} = \frac{1}{4\epsilon}\,\delta''(R) +
  O(\epsilon^0)\;,
\end{equation}
we find
\begin{widetext}
\begin{equation}
  \label{eq:poleterm}
   {\left[ \partial_{\mathring{w}} \widetilde{G^3}
     \right]}_{\mathring{w}=0}
  = \frac{-3}{4\epsilon}\,S_{\bar{m}}\frac{\partial^2}{\partial R^2}\,
 \overline{e^{iR \bm{p}{\cdot}\bm{e}}}^{\bar{m}}
{\int}d^m\upsilon\,e^{i\sqrt{R}\,\bm{k}{\cdot}\bm{\upsilon}}\,
\Phi_{m,d*}^2(\upsilon)\, \sum_{\alpha}\partial_{\upsilon_{\alpha}}^4\,
Y_{m,d^*}{\big(\upsilon^2\big)}\bigg|_{R=0} + O(\epsilon^0)\;.
\end{equation}

The right-most part of Eq.~(\ref{eq:poleterm}) can be rewritten as
\begin{eqnarray}
  \label{eq:derterm}
\sum_{\alpha=1}^m\partial_{\upsilon_{\alpha}}^4\,
Y_{m,d^*}{\big(\upsilon^2\big)}
&=&12m\,Y_{m,d^*}''{\big(\upsilon^2\big)}
+48\upsilon^2\,Y_{m,d^*}'''{\big(\upsilon^2\big)}+
16\sum_{\alpha}\upsilon_{\alpha}^4\,
Y_{m,d^*}^{(iv)}{\big(\upsilon^2\big)}\;,
\end{eqnarray}
where $Y^{(iv)}(z) \equiv d^4Y(z)/dz^4$.
\end{widetext}

Note that the function of $R$ that must be differentiated in
Eq.~(\ref{eq:poleterm}) has an expansion in even powers of $\sqrt{R}$
because the angular integration in $\int d^m\upsilon$ yields zero for
the coefficients of all odd powers. Thus only the contributions
$\propto R^2$ produced by the two exponential functions contribute to
the pole term on the right-hand side.

For the angular averages, we may exploit the characteristic function
\begin{equation}
  \label{eq:genfun}
  \overline{e^{i\bm{k}{\cdot}{\bm{\upsilon}}}}^m
 =\sum_{{\ell}=0}^\infty\frac{(ik \upsilon)^{2{\ell}}\,
   \Gamma{\big(\frac{m}{2}\big)}}{4^{\ell}\,{\ell}!\,
   \Gamma{\big({\ell}+\frac{m}{2}\big)}}
\end{equation}
to find
\begin{eqnarray}
       \label{eq:prodangav}
     \lefteqn{\overline{\upsilon_{\alpha_1}\cdots
         \upsilon_{\alpha_{2N}}}^{m}/ \upsilon^{2N}} &&\nonumber\\
&=& \frac{\delta_{\alpha_1\alpha_2}
  \cdots\delta_{\alpha_{2N-1}\alpha_{2N}}+
  \ldots}{m(m+2)(m+4)\cdots[m+2(N-1)]}\;.
\end{eqnarray}
In the numerator the ellipsis $\ldots$ stands for the remaining
${(2N-1)!!-1}$ pairings of the indices. This can also be thought of as
$2N$-point function of a Gaussian theory whose propagator between two
``points'' $\alpha_i$ and $\alpha_j$ is given by
$\delta_{\alpha_i\alpha_j}$.

Utilizing these results, we easily find
\begin{equation}
  \label{eq:angaveq:angavalphasyres}
  \frac{\partial^2}{\partial R^2} \,
  \overline{e^{iR\bm{p}{\cdot}{\bm{e}}}}^{\bar{m}}
      \Big|_{R=0}=-\frac{p^2}{\bar{m}}\;,
     \end{equation}
\begin{equation}
  \label{eq:angavu4}
       \sum_{\alpha}\overline{\upsilon_{\alpha}^4}^{m}
       = \frac{3\,\upsilon^4}{m+2}\;,
\end{equation}
\begin{equation}
  \label{eq:angavalphasym}
 \frac{\partial^2}{\partial R^2}\,
\overline{e^{i\sqrt{R}\,\bm{k}{\cdot}\bm{\upsilon}}}^{m}\Big|_{R=0}
     =  \frac{k^4 \, \upsilon^4}{4m(m+2)}\;,
\end{equation}
and
\begin{widetext}
\begin{eqnarray}
  \label{eq:angavalphacub}
\frac{\partial^2}{\partial R^2}
  \sum_{\alpha}\overline{\upsilon_{\alpha}^4\, e^{i\sqrt{R}\,
    \bm{k}{\cdot}\bm{\upsilon}}}^{\Omega_m}\Big|_{R=0}
     &=&\frac{3(m+8)\,k^4 + \check{k}^4}{4m(m+2)(m+4)(m+6)}\,\upsilon^8\;.
\end{eqnarray}

Combined with Eq.~(\ref{eq:poleterm}), these results yield
\begin{equation}
  \label{eq:G3wres}
{\left[ \partial_{\mathring{w}} \widetilde{G^3} \right]}_{\mathring{w}=0}=
  \frac{F_{m,\epsilon }^2}{\epsilon}
\left[ \frac{18\,i_\phi (m)}{8-m}\,p^2 -
  \frac{9\,i_{\sigma_1}(m)}{4m(m+2)}k^4
- \frac{24\,i_{\sigma_2}(m)}{m(m+2)(m+4)(m+6)}\,\check{k}^4\right]
+ O(\epsilon^0)
\end{equation}
where we have introduced the integrals
\begin{equation}
  \label{eq:iphidef}
  i_\phi(m)\equiv B_m{\int_0^\infty}d\upsilon\, \upsilon^{m-1}\,
  \Phi^2_{m,d^*}(\upsilon) \,{\left[m\,Y_{m,d^*}''{\big(\upsilon^2\big)}+
      4\upsilon^2 \,Y_{m,d^*}'''{\big(\upsilon^2\big)}
 +\frac{4}{m+2}\,\upsilon^4\,
 Y_{m,d^*}^{(iv)}{\big(\upsilon^2\big)}\right]}\;,
\end{equation}
\begin{equation}
  \label{eq:isigma1def}
  i_{\sigma_1}(m)\equiv B_m\,{\int_0^\infty}d\upsilon\, \upsilon^{m+3}
{\left[m  
Y_{m,d^*}''{\big(\upsilon^2\big)}+4\,\upsilon^2\,Y_{m,d^*}'''{\big(\upsilon^2\big)}
 + \frac{4\,(m+8)\,\upsilon^4}{(m+4)(m+6)}\,
 Y_{m,d^*}^{(iv)}{\big(\upsilon^2\big)} \right]}\;,
\end{equation}
and
\begin{equation}
  \label{eq:isigma2def}
  i_{\sigma_2}(m)\equiv B_m\,{\int_0^\infty}d\upsilon\, \upsilon^{m+7}\,
  \Phi^2_{m,d^*}(\upsilon)\,Y_{m,d^*}^{(iv)}{\big(\upsilon^2\big)}\;,
\end{equation}
with
\begin{equation}
  \label{eq:Bm}
  B_m\equiv \frac{S_{4-{m/2}}\, S_m}{F_{m,0}^2}=
{\frac{{2^{10 + m}}\,{{\pi }^{6 + {{3\,m}/{4}}}}\,
     \Gamma({{m}/{2}})}{\Gamma(2 - {m}/{4})\,
     {{\Gamma^2({m}/{4}})}}}\;.
\end{equation}
For completeness, we recall the definitions\cite{SD01} of
\begin{equation}\label{eq:jphidef}
j_\phi(m)\equiv B_m \,
{\int_0^\infty}\!{d}\upsilon\,\upsilon^{m-1}\,
\Phi_{m,d^*}^3(\upsilon)\;,
\end{equation}
and
\begin{equation}
  \label{eq:jsigmadef}
j_\sigma(m)\equiv B_m\,{\int_0^\infty}\!{d}\upsilon\,\upsilon^{m+3}\,
\Phi_{m,d^*}^3(\upsilon)\;.
\end{equation}

\section{The scaling functions $\Phi_{m,d}(\upsilon)$ and $Y_{m,d}(z)$}
\label{sec:scfY}

First, let us recall the properties of the scaling function
$\Phi_{m,d}(\upsilon)$, which were established in
Refs.~\onlinecite{DS00a} and \onlinecite{SD01}. Its Taylor expansion
is respectively,
\begin{equation}
\label{eq:Ffmd}
\Phi_{m,d}(\upsilon)=\frac{\pi^{(1-d)/2}}{2^{2+m}}\,
\sum_{\ell =0}^\infty \frac{\Gamma {\big( \frac {\ell}{2} +1 -
\frac{\epsilon}{2} \big)}}
{\ell{!}\,\Gamma{\left(\frac {\ell}{2}+ \frac {1}{2}+ \frac {m}{4}
    \right)}}
\bigg(-\frac {\upsilon^2}{4}\bigg)^\ell \;.
\end{equation}
It is possible to express the power series in closed form by
exploiting a relation of the generalized hypergeometric functions
${_1\!}F_2$ (by summing the contributions with even and odd values of
$ \ell $ separately), namely,
\begin{equation}
\label{eq:sum1F2}
\sum_{\ell =0}^\infty\frac{1}{\ell!}\,
\frac{\Gamma{\big(a+\frac{\ell}{2}\big)}}
{\Gamma{\big(b+\frac{\ell }{2}\big)}}\left(-y \right)^\ell
=\frac{\Gamma(a)}{\Gamma(b)}
\;{{_{1\!}F_{2}}}{\bigg(a;\frac{1}{2},b;\frac{y^2}{4}\bigg)} -y\;
\frac{\Gamma{\big(a+\frac{1}{2}\big)}}{\Gamma{\big(b+\frac{1}{2}\big)}}
\; {{_{1\!}F_{2}}}{\bigg(a+\frac{1}{2};\frac{3}{2},
  b+\frac{1}{2};\frac{y^2}{4}\bigg)}\,.
\end{equation}
The large $\upsilon$~properties of ${_1\!}F_2$ lead us to the
asymptotic expansion
\begin{equation}
\label{eq:asexpPhi}
\Phi_{m,d^{*}}(\upsilon)\mathop{\approx}\limits_{\upsilon\to \infty }
\frac{2^{1-m}}{\pi^{(6+m)/4}}\,
\frac{m-2}{\Gamma {\big(\frac 12+\frac m4\big)}}\,
\upsilon^{-4}\, {\Big[1+O{\big(\upsilon^{-4}\big)}\Big]}\;,
\end{equation}
which will be used in numerical evaluations of $i_{\sigma_2}(m)$
described in Sec.~\ref{sec:estim}.

The scaling function $Y_{m,d}(\upsilon^2)$ was also studied in
Ref.~\onlinecite{SD01}: It appeared in the calculation of another
scaling function, $\Theta(\upsilon)$, that was needed for the coupling
constant renormalization at the two-loop level. Its Taylor expansion,
from Eq.~(D4) in this reference, reads
\begin{equation}
\label{eq:Ymd}
Y_{m,d}(z)=\,\frac{\pi^{(1-d)/2}}{2^{4+m}}\,\sum_{\ell =0}^\infty
\frac {\Gamma {\big(\frac{\ell}{2}-\frac \epsilon 2\big)}}
      {\ell!\,\Gamma {\left(\frac{\ell}{2}+\frac{1}{2}+\frac{m}{4}
          \right)}}
\Big(-\frac{z}{4}\Big)^\ell \;.
\end{equation}
As for $\Phi $ above, we can write a closed form for this function in
terms of two hypergeometric functions:
\begin{equation}
\label{eq:yFPQ}
Y_{m,d}(z)=\frac{\pi^{(1-d)/2}}{2^{4+m}}\,
{\bigg[\frac{\Gamma{\big(-\frac{\epsilon}{2}
      \big)}}{\Gamma{\big(\frac{1}{2}+\frac{m}{4}\big)}}\,
{_{1\!}}F_2{\bigg(-\frac{\epsilon}{2};\frac{1}{2}+\frac{m}{4};
  \frac{z^2}{64}\bigg)}- \frac{z}{4}\,
\frac{\Gamma {\big(\frac{1}{2}- \frac{\epsilon}{2}\big)}}{\Gamma
  {\big(1+\frac{m}{4}\big)}}\,
{_{1\!}}F_2{\bigg( \frac 12-\frac{\epsilon}{2};
  \frac{3}{2},1+\frac{m}{4};\frac{z^2}{64}\bigg)}\bigg]}\;.
\end{equation}

Let us note that, though the series~(\ref{eq:Ymd}) starts with a
term which has a pole in $\epsilon $, all its derivatives are regular
at $\epsilon =0$. In particular, for the calculation of the integral
$i_{\sigma_2}(m)$, we need only $Y_{m,d^{*}}^{(iv)}(z)$ at the upper
critical dimension $d^{*}=4+m/2$. Thus,
\begin{eqnarray}
\label{eq:Yiv}
\lefteqn{%
Y_{m,d^*}^{(iv)}(z)=
 \frac{1}{2^{12+m}\,\pi^{(6+m)/4}}\,
\sum_{\ell =0}^\infty\frac{1}{\ell!}\,
\frac{\Gamma{\big(2+\frac{\ell }{2}\big)}\,({-}\frac{z}{4})^\ell }
  {\Gamma{\big(\frac{5}{2}+\frac{m}{4} +\frac{\ell }{2}\big)}}
}&&\\
&=&\frac{1}{2^{14+m}\,\pi^{(6+m)/4}}\,
\left[ \frac{4}{\Gamma{\big(\frac{5}{2}+\frac{m}{4}\big)}}\;
{{_{1\!}F_{2}}}\!\left(2;\frac{1}{2},\frac{5}{2}+\frac{m}{4};
  \frac{z^2}{64}\right)
- \frac{z\, \Gamma{\big(\frac{5}{2}\big)}}{
  \Gamma{\big(3+\frac{m}{4}\big)}} \;
{{_{1\!}F_{2}}}\!\bigg(\frac{5}{2};\frac{3}{2},3+\frac{m}{4};
\frac{z^2}{64}\bigg)
\right]\,.
\end{eqnarray}

In the special cases $m=2$ and $m=6$, the hypergeometric functions here reduce to
the elementary functions:
\begin{equation}
  \label{eq:Yiv2}
  Y_{2,5}^{(iv)}(z)=\frac{3}{(4\,\pi)^2\,z^4}\,{\left[1
      -e^{-{z}/{4}}\,{\left(1 + \frac{z}{4} +
          \frac{z^2}{32} +
          \frac{z^3}{384}\right)}\right]}
\end{equation}
and
\begin{equation}
  \label{eq:Yiv6}
 Y_{6,7}^{(iv)}(z)=
  \frac{48\,\left( -320 + z^2 \right)  + e^{-z/4}
    {\left[15360 + z\,\big( 16 + z \big)
        \big( 240 + 12\,z + z^2
        \big)\right]}}{(4\,\pi)^3\,64\,z^{6}}\;.
\end{equation}
\end{widetext}

For the integral $i_{\sigma_2}(m)$, we will need the asymptotic
expansion of $Y^{(iv)}_{m,d^*}(z)$:
\begin{eqnarray}
  \label{eq:asexpYiv}
 && Y^{(iv)}_{m,d^*}(z)\mathop{\approx}\limits_{z\to\infty}\\
&&\nonumber
\frac{3}{2^{2+m}\,\pi^{(6+m)/4}\,
  \Gamma{\big(\frac{1}{2}+\frac{m}{4}\big)}}\,
z^{-4}\,{\Big[1+O{\big(z^{-2}\big)}\Big]}\;.
\end{eqnarray}

Let us however note that, for arbitrary values of $m$, derivatives of
$Y_{m,d^*}(z)$ can be expressed completely in terms of
$\Phi_{m,d^*}(\upsilon)$ (with $\upsilon=\sqrt{z}$),\cite{DS00a,SD01}
since their Taylor series expansions differ only slightly. In
particular, using
\begin{equation}
\Gamma{\Big( \frac {\ell}{2} +1 -\frac {\epsilon}{2} \Big)} =
{\Big( \frac {\ell}{2}-\frac {\epsilon}{2} \Big)}
\,\Gamma{\Big( \frac {\ell}{2} -\frac {\epsilon}{2} \Big)}\,,
\end{equation}
we get the general relation
\begin{equation}
\label{Gryf}
\Phi_{m,d}(\sqrt{z}) = 2 \big[ z\,Y'_{m,d}(z) -
\epsilon\,Y_{m,d}(z)\big]\,.
\end{equation}
Recalling that $Y_{m,d}$ has a pole at $\epsilon =0$, we verify that
$Y_{m,d}(z)=-\Phi_{m,d^*}(0)\,/(2\epsilon) +  O(\epsilon^0)$.
Thus, in the limit $\epsilon\to 0$, we find an exceedingly simple
relationship between $Y'_{m,d}$ and $\Phi_{m,d}$:
\begin{equation}
\label{eq:Grel}
Y'_{m,d^*}(z)= \frac{1}{2z} \Big[\Phi_{m,d^*}(\sqrt{z})-
\Phi_{m,d^*}(0)\Big]\,.
\end{equation}
In other words, the scaling function $\Phi_{m,d^*}$ completely
determines all derivatives of $Y_{m,d^*}$, a fact we will exploit in the
following Appendix.

\section{Analytic results for $m=2$ and $m=6$}
\label{sec:m2m6}

The special case $m=2$ is the simplest one since the scaling function
$\Phi$ acquires an extremely simple form at the upper critical
dimension:
\begin{equation}\label{fidva}
\Phi_{2,5}(\sqrt{z})=\frac{1}{(4 \pi)^2}\, e^{-z/4}\,.
\end{equation}

For the function $Y'_{2,5}$, using the relation (\ref{eq:Grel}), we
thus get
\begin{equation}
Y'_{2,5}(z)=\frac{1}{(4 \pi)^2}\,\frac{1}{2z}\left(e^{-z/4}-1\right)\,.
\end{equation}
Even more simplifications of integrals are realised if we use the
representation
\begin{equation}\label{Irit}
\int_0^1dt\,e^{-zt/4}=\,\frac{4}{z}(1-e^{-z/4})\,,
\end{equation}
so that
\begin{equation}
Y_{2,5}^{(iv)}(z)=\frac{1}{(4\pi)^2}\,\frac{2}{4^5}\,
\int_0^1dt\,t^3 e^{-zt/4}\,.
\end{equation}

Inserting this expression into Eq.~(\ref{eq:isigma2def}) for the
integral $i_{\sigma_2}$, we find
\begin{equation}
i_{\sigma_2}(2)=2 \int_0^\infty\!\! d\,\zeta \,\zeta^4\,e^{-2\zeta}
\int_0^1dt\,t^3 e^{-\zeta t}\,.
\end{equation}
To arrive here, we changed the integration variable to $\zeta \equiv
z/4$ for convenience. Carrying out the trivial integrations, we get
\begin{equation}
i_{\sigma_2}(2)=2\cdot 4! \int_0^1dt\,t^3(t+2)^{-5}=\frac{2}{27}\,.
\end{equation}

The special case $m=6$ is, as usual, somewhat more involved. This is
due to a more complicated functional form of the corresponding scaling
function $\Phi$.  Using $\zeta =z/4$ once more, we have
\begin{equation}\label{eq:Phidstar6x}
\Phi_{6,7}{\big(\sqrt{4\zeta}\big)} = \frac{1}{2(4
  \pi)^3}\,\frac{1}{\zeta^2}
\Big[1-(1+\zeta) e^{-\zeta}\Big]\,.
\end{equation}
Again, we exploit an integral representation like Eq.~(\ref{Irit}) ---
now, for both functions $\Phi_{6,7}$ and $Y_{6,7}'$. Thus,
(\ref{eq:Phidstar6x}) can be written as
\begin{equation}\label{phint}
\Phi_{6,7}{\big(\sqrt{4\zeta}\big)}
=\frac{1}{2\,(4\pi)^3}\int_0^1dt\,t\,e^{-\zeta t}\,.
\end{equation}
Inserting this into the relation (\ref{eq:Grel}), we obtain
\begin{equation}
Y_{6,7}'(z)=-\frac{1}{16\,(4\pi)^3}\int_0^1dt\,t^2\!\!\int_0^1dy\,
e^{-z ty/4}\,,
\end{equation}
and
\begin{equation}
Y_{6,7}^{(iv)}(z)= \frac{1}{4^5\,(4\pi)^3}
\int_0^1dt\,t^5\!\!\int_0^1dy\,y^3\,e^{-z ty/4}\,.
\end{equation}
The desired integral $i_{\sigma_2}(6)$ thus becomes
\begin{widetext}
\begin{eqnarray}
i_{\sigma_2}(6)&=&4
\int_0^\infty\! d\zeta\, {\zeta}^6 \!\!
\int_0^1\!d\alpha\,\alpha\,e^{-\alpha \zeta} \!\!
\int_0^1\!d\beta\,\beta\,e^{-\beta \zeta}
\int_0^1\!dt\,t^5\!\!
\int_0^1\!dy\,y^3\,e^{-\zeta ty}
\nonumber\\
&=&4\cdot 6!
\int_0^1\!d\alpha\,\alpha\int_0^1\!d\beta\,\beta
\int_0^1\!dt\,t^5\!\int_0^1\!dy\,y^3\,\frac{1}{(\alpha+\beta+ty)^7}
=20{\Big(8\,\ln\frac{4}{3}-\frac{59}{27}\Big)}\,.
\end{eqnarray}
\end{widetext}
The integration over $\zeta$ was trivial, and the last result was
obtained by simple repeated integrations with the help of {\sc
  Mathematica}.\cite{rem:MAT}


\begin{thebibliography}{33}
\bibitem[{rem({\natexlab{b}})}]{rem:elcoeff} \bibinfo{note}{There are
    fewer tensors ${\mathcal T}_{\alpha_1\alpha_2\alpha_3\alpha_4}$
    than elastic moduli $C_{\alpha_1\alpha_2;\alpha_3\alpha_4}$
    because the latter are not totally symmetric but satisfy in
    general only the relations\cite{LL83}
    $C_{\alpha_1\alpha_2;\alpha_3\alpha_4}=
    C_{\alpha_2\alpha_1;\alpha_3\alpha_4} =
    C_{\alpha_1\alpha_2;\alpha_4\alpha_3} =
  C_{\alpha_3\alpha_4;\alpha_1\alpha_2}$. 
  coefficien$\mathring{\sigma}_i$ than 
  $C_{\alpha_1\alpha_2;\alpha_3\alpha_4}$ because the 
  symmetric but satisfy in 
  $C_{\alpha_1\alpha_2;\alpha_3\alpha_4}= %
  C_{\alpha_2\alpha_1;\alpha_3\alpha_4} = %
  C_{\alpha_1\alpha_2;\alpha_4\alpha_3} = %
  C_{\alpha_3\alpha_4;\alpha_1\alpha_2}$.}
\bibitem[{com()}]{com:aniso} \bibinfo{note}{To investigate the
    Lifshitz-point critical behavior in the spherical limit
    $n\to\infty$, Frachebourg and Henkel\cite{FH93} considered a
    Hamiltonian of the form (\ref{eq:Hamaniso}) but without the
    isotropic derivative term $\propto\mathring{\sigma}_1$. In RG
    analyses such a simplifying assumption cannot be maintained
    because this term gets generated under the RG, even within the
    framework of the $1/n$ expansion.}
\bibitem[{rem({\natexlab{d}})}]{rem:accuracy} \bibinfo{note}{We have
    verified that our numerical integration method gives values for
    $m=2$ and $m=6$ in conformity with the analytical results
    (\ref{eq:jsigmam26}). Yet, in the $m=6$ case, we have contented
    ourselves with a numerical accuracy $\lesssim 0.4$ \%. The
    additional decimals given in Table~\ref{tab:jis} were determined
    from the exact expression (\ref{eq:jsigmam26}). For $m=2$, the
    numerical accuracy is higher, corresponding essentially to the
    number of decimals given.}
\end{thebibliography}

\end{document}